\documentclass[onecolumn]{aa} 
\usepackage{graphicx}
\usepackage{natbib}
\usepackage{deluxetable}
\usepackage{url}
\usepackage{txfonts}
\begin{document}
   \title{Multi-Conjugate Adaptive Optics images of the Trapezium Cluster\thanks{Observations made at the ESO Paranal Observatory}}

   \author{H. Bouy\inst{1,2} \thanks{Marie Curie Outgoing International Fellow MOIF-CT-2005-8389}
          \and J. Kolb\inst{3}
          \and E. Marchetti\inst{3}
	  \and E.~L. Mart\'\i n\inst{2,4}
          \and N. Hu\'elamo\inst{5}
 	  \and D. Barrado y Navascu\'es\inst{5}
          }

   \offprints{H. Bouy}

   \institute{Astronomy Department, University of California, Berkeley, CA 94720, USA\\
     \email{hbouy@astro.berkeley.edu}
     \and
     Instituto de Astrof\'\i sica de Canarias, C/ V\'\i a L\'actea s/n, E-38200 - La Laguna, Tenerife, Spain\\
     \email{ege@iac.es}
     \and
     European Southern Observatory, Karl Schwartzschild Str. 2, D-85748 Garching bei M\"unchen, Germany\\
     \email{jkolb@eso.org, emarchet@eso.org}
     \and
     University of Central Florida, Department of Physics, P.O. Box 162385, Orlando, FL 32816-2385, USA
     \and
     Laboratorio de Astrof\'\i sca Espacial y F\'\i sica Fundamental (LAEFF-INTA), Apdo 50727, 28080 Madrid, Spain\\
     \email{nhuelamo@laeff.inta.es,barrado@laeff.inta.es}
   }

   \date{accepted September 25, 2007}

 
  \abstract
   {{\bf High resolution version of the manuscript available of http://arrakeen.free.fr/pub/madorion.pdf}. Multi-Conjugate Adaptive Optics (MCAO) combines the advantages of standard adaptive optics, which provides high contrast and high spatial resolution, and of wide field ($\approx$1\arcmin) imaging. Up to recently, MCAO for astronomy was limited to laboratory experiments. In this paper, we present the first scientific results obtained with the first MCAO instrument put on the sky.}
   {We present a new study of the Trapezium cluster using deep MCAO images with a field of view of 1\arcmin$\times$1\arcmin obtained at the VLT.}
   {We have used deep J, H and Ks images recently obtained with the prototype MCAO facility MAD at the VLT in order to search for new members and new multiple systems in the Trapezium cluster. On bright targets (Ks$\approx$9~mag), these images allow us to reach $\Delta$Ks$\approx$6~mag as close as 0\farcs4.}
   {We report the detection of 128 sources, including 10 new faint objects in the magnitude range between 16.1$<$Ks$<$17.9~mag. In addition to all previously known multiple systems with separations greater than 0\farcs1, we confirm the multiplicity of TCC-055. We also report the detection in J, H and Ks of a very red extended embedded protostellar object, HC~419, previously detected in the thermal infrared only.}
   {The analysis of the first MCAO images obtained on the sky demonstrates not only the technical feasibility of MCAO but also its great potential and versatility in terms of scientific outputs.}

   \keywords{Instrumentation: Adaptive optics, Techniques: High Angular Resolution, Stars: visual binaries, Stars: Evolution, Stars: formation, Stars: general}

   \maketitle
%

\section{Introduction}

Since the demonstration of the capabilities of Adaptive Optics for Astronomy in the early 90's \citep[see e.g][and references therein]{1991SPIE.1542..248R, 1993Icar..105..263S}, a large number of observatories have made the choice of equipping their telescopes with such facilities. Back then, only $\approx$2\% of the sky was accessible to adaptive optics instruments, because of the need for a sufficiently bright star nearby the target of interest. The development of artificial Laser Guide Star \citep[LGS, ][]{1995SPIE.2534..150F} and of near-IR wavefront sensors such as NACO at the VLT \citep{2002Msngr.107....1B} allowed to close the loop on fainter objects. While these technology increased the sky coverage, the corrected field of view (FoV) of the adaptive optics was still limited because of the typical scale of the atmospheric perturbations (10--20\arcsec\, at near-infrared wavelengths). This limitation was foreseen even before the first adaptive optics instrument for astronomy had been built. In his review, \citet{1988vltt.conf..693B} first proposed the development of Multi-Conjugate Adaptive Optics (MCAO), aiming at enlarging the size of the FoV corrected for atmospheric turbulence. The principle of MCAO is based on probing the volume of atmospheric turbulence above the telescope by performing wavefront sensing on not only one but several guide stars in the FoV, implementing a tomographic reconstruction of the turbulence \citep{2000Natur.403...54R} to reveal its 3-D distribution and then applying the correction in the FoV by mean of several deformable mirrors optically conjugated at different altitudes in the atmosphere. In this way the correction spreads efficiently in the FoV and it is not concentrated only for a specified direction as in classical AO system.

In this paper, we present the analysis of multi-conjugate adaptive optics images of the Trapezium cluster obtained at the VLT with the \emph{Multi-Conjugate Adaptive Optics Demonstrator} (hereafter MAD). These images reach an unprecedented depth on large field of view with a resolution equivalent or better than the resolution of HST equivalent instrument NICMOS3. After describing the instrument in Section 2, the observations of the Trapezium cluster in Section 3, and the performances of the instrument in Section 4, we discuss the scientific results in Sections 5, 6 and 7.
 
\section{MAD: a multi-conjugate adaptive optics facility at the VLT}
MAD is a prototype instrument performing wide field-of-view real-time correction for atmospheric turbulence \citep{2006SPIE.6272E..21M}. MAD has been built by the European Southern Observatory (ESO) with the contribution of two external consortia to prove on the sky the feasibility of Multi-Conjugate Adaptive Optics technique in the framework of the 2$^{\rm nd}$ generation VLT instrumentation and of the European Extremely Large Telescope \citep[ELT, ][]{2007Msngr.127...11G}. 

MAD is equipped with three Shack-Hartmann wavefront sensors \citep{1900ApJ....11..400H,1971JOSA....61...656} for measuring the atmospheric turbulence from three guide stars located in a circular FoV of 2\arcmin. Each wavefront sensor can patrol the FoV to acquire signal for any geometry of guide stars surrounding the astronomical target. Depending on the atmospheric seeing conditions the limiting magnitude for the guide stars can be up to V$\approx$13~mag. Two 60 elements bimorph deformable mirrors optically conjugated at 0 and 8.5~km in the atmosphere above the telescope ensure the MCAO correction. A dichroic located between the deformable mirrors and the wavefront sensors reflects the visible light toward the latter ones and transmits the infrared light to CAMCAO (CAmera for MCAO), the MAD infrared scientific imaging camera. CAMCAO is based on a 2048$\times$2048 pixels HAWAII-2 infrared detector with a pixel scale of 0\farcs028 for a total FoV of 57\farcs3. It is mounted on a movable table to scan the full 2\arcmin\, FoV. This scanning capability allows to implement efficient dithering for infrared sky subtraction without modifying the telescope pointing and maintaining the adaptive optics correction loop closed during the full observation. The camera is equipped with a standard set of J, H and Ks filters plus two additional narrow band filters (Br$\gamma$ 2.166~$\mu$m and Br$\gamma$ continuum 2.144~$\mu$m). Originally designed as a laboratory experiment, MAD has been installed at the Visitor Focus of the VLT telescope UT3 (Melipal) located at the ESO Paranal Observatory on February 2007 and performed the first on-sky demonstration run from March 25$^{\rm th}$ to April 6$^{\rm th}$. 

\section{Observations and data reduction}
\begin{figure*}
   \centering
   \includegraphics[width=0.55\textwidth]{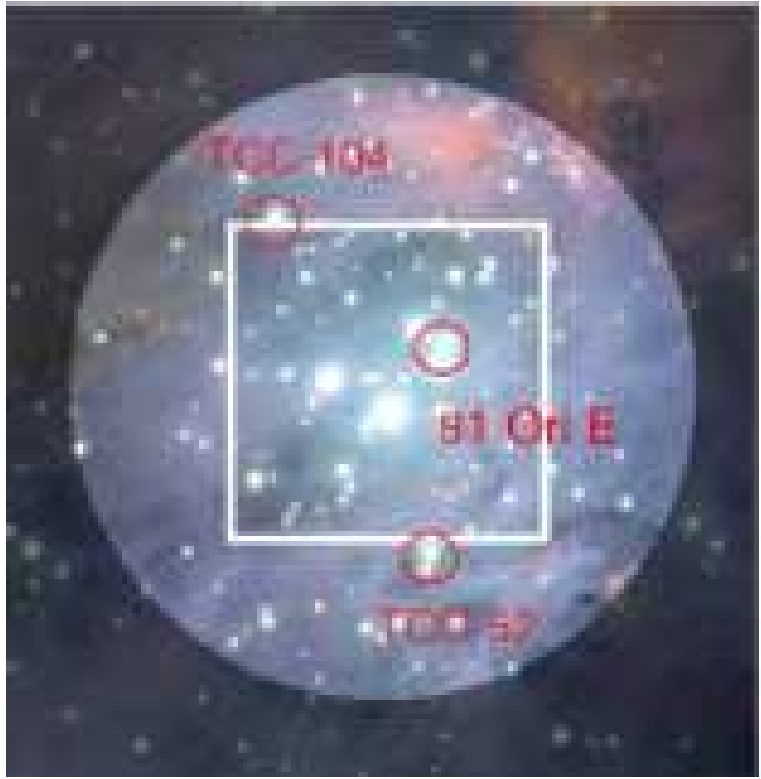}
      \caption{ISAAC three color image (Js, H, Ks, Credit Mark McCaughrean and the European Southern Observatory) showing the circular field of view (1\arcmin\, radius) of the MAD wavefront sensor, and the 1\arcmin$\times$1\arcmin box field of view of the CAMCAO scientific camera. The three stars used for wavefront sensing are indicated with red circles. North is up and east is left}
         \label{2mass}
   \end{figure*}

During the first on-sky demonstration run of MAD a region of ~1\arcmin$\times$1\arcmin\, around the Trapezium cluster in the Orion nebula M42 has been observed. The field is centered on $\alpha$=05h35min16.5sec, $\delta$=-05\degr23\arcmin13.47\arcsec\, (J2000). The wavefront sensing was performed using three guide stars, namely $\theta^{1}$~Ori~E (V=11.1~mag), TCC-053 (V=12.5~mag) and TCC-104 (V=11.2~mag). Figure \ref{2mass}  shows the fields of view of MAD, of CAMCAO and the 3 guide stars. The geometrical distribution of the guide stars is quite asymmetric, but it has been found to be the only one allowing to cover a reasonable FoV and having stars bright enough at visible wavelengths to operate a reasonably efficient wavefront sensing. The MCAO loop has been closed at a correction frequency of 400~Hz.

All the observations have been carried out on April 4$^{\rm th}$ in the early evening due to the low elevation of the target (1.4$<$airmass$<$1.7), as part of the first technical demonstration run. The publicly available data have been downloaded from the ESO public archive\footnote{The raw datasets (ID MAD\_TR.2007-05-11) are available at \url{http://www.eso.org/projects/aot/mad/commdata/}. The processed images can be requested to the authors.}. For each band a set of 25 images of 15$\times$0.79~s (\verb|NDIT|$\times$\verb|DIT|) have been retrieved by dithering on a box of 20\arcsec\, using the scanning capability of the infrared camera and keeping the adaptive optics loop closed during the whole operations. The average seeing\footnote{at the zenith and in the visible} during the observations, as reported by the ESO Ambient Conditions Database, was 1\farcs0.

\begin{figure*}
   \centering
   \includegraphics[width=0.85\textwidth]{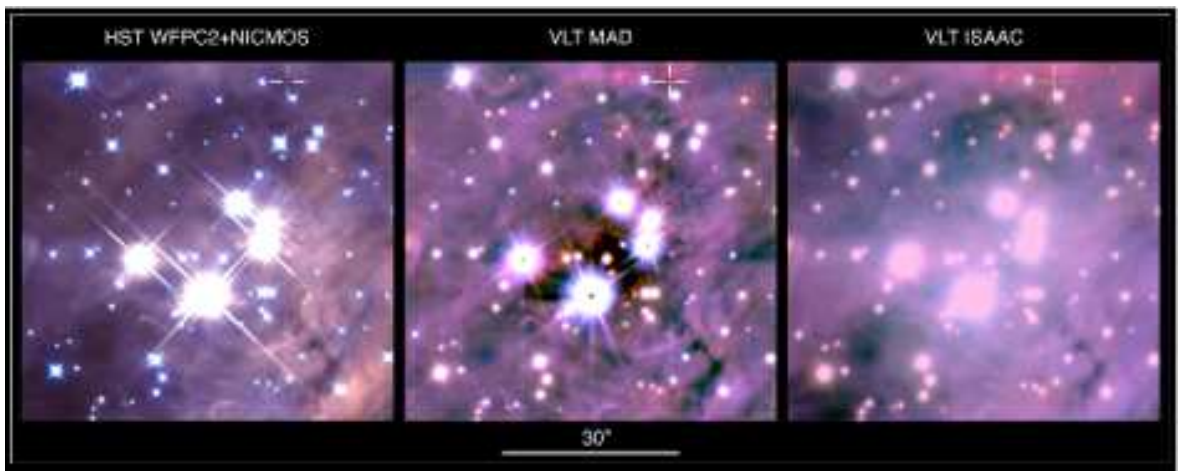}
      \caption{Color images of the Trapezium obtained with different instruments. \emph{Left:} HST NICMOS+WFPC2 (F547M, F110W and F160W, Credit O'Dell and STSci); \emph{Center:} VLT MAD (J, H and Ks); \emph{Right:} VLT ISAAC (Js, H, Ks, Credit Mark McCaughrean and ESO). North is up and east is left. The scale is indicated. A cross indicates the location of HC~419. }
         \label{orion_mad_hst}
   \end{figure*}

The Trapezium cluster was observed in the J, H, Ks and Br$\gamma$ (2.166~$\mu$m) filters. The corresponding images have been dark subtracted, flat-fielded, sky subtracted and stacked using standard procedures with the \emph{Eclipse} reduction package \citep{1997Msngr..87}. The astrometric solution was computed using isolated and unresolved 2MASS counterparts. Figure \ref{orion_mad_hst} shows the final image and compare to HST WFPC2+NICMOS and VLT ISAAC images. No Br$\gamma$  continuum images were obtained. The final processed mosaics are made available upon request to the authors of this article.

\section{Performances and data analysis}

MAD at the VLT is a first-of-its-kind wide field adaptive optics instrument. The analysis of this new kind of data requires special attention. The Trapezium cluster being a crowded field and the extinction being strong and spatially variable, we opted for PSF photometry rather than aperture photometry. Because of the complexity of the MCAO wavefront sensing, the PSF shows space variations due to anisoplanatic effects in the AO observations that can affect the PSF photometry. In order to minimize these effects, we used the \emph{daophot} package within IRAF to extract the list of sources and perform PSF photometry using a second order polynomial variable PSF. A more detailed analysis of the photometry in MCAO images will be presented in a coming paper (Bouy, Marchetti \& Kolb, in prep.). The PSF is well sampled only in the Ks band. The J and H band fluxes are therefore less reliable. We used well behaved isolated and unresolved COUP sources with ISAAC Js, H and Ks photometry \citep{2005ApJS..160..319G} to derive the following photometric instrumental zeropoints: ZP(J)=24.7~mag, ZP(H)=24.4~mag, ZP(Ks)=23.6~mag (including exposure times). From the standard deviation of the five zeropoint reference stars, we estimate the final uncertainties to be 0.2, 0.3 and 0.3~mag in the Ks, H and J band respectively. As a sanity check, we also compare with the zeropoints obtained  using the few isolated and well behaved 2MASS sources (with a quality flag equal to \verb|AAA|) present in the field of view of the MAD images. The corresponding zeropoints agree well with the previous ones within the uncertainties. For the common sources, the final photometry agrees well with the COUP and \citet{2004AJ....128.1254L} photometry within the uncertainties.

\subsection{MCAO correction}
\begin{figure*}
   \centering
   \includegraphics[width=0.95\textwidth]{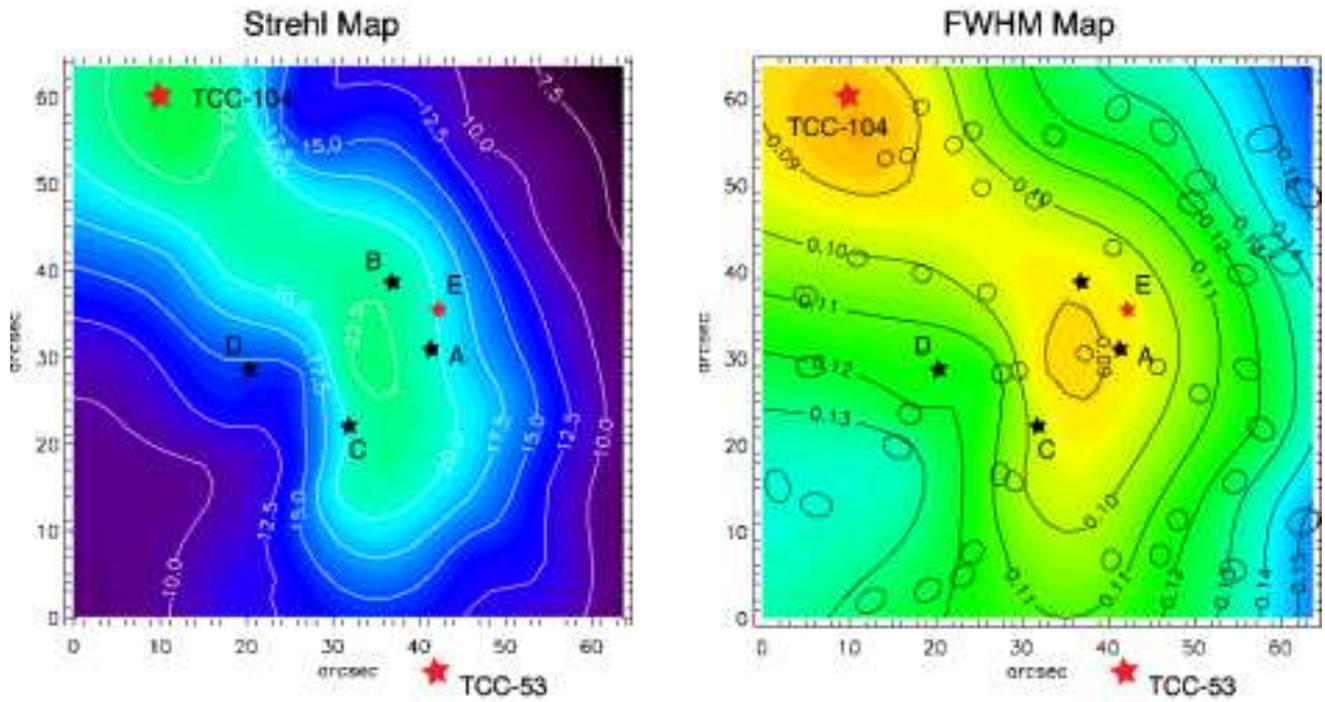}
      \caption{Strehl map (left) and FWHM map (right) the MAD Ks image. North is up and east is left. Ellipses with semi-major axis and orientations representative of the measured values for the PSF of the stars used for the computation are also over-plotted. The Trapezium $\theta$~Ori A, B, C, D stars are indicated (black stars), as well as the three reference stars used for wavefront sensing (TCC-104, TCC-053 and $\theta$~Ori E, red stars). The scale is indicated with contour plots.}
         \label{strehlmap}
   \end{figure*}

\begin{figure*}
   \centering
   \includegraphics[width=\textwidth]{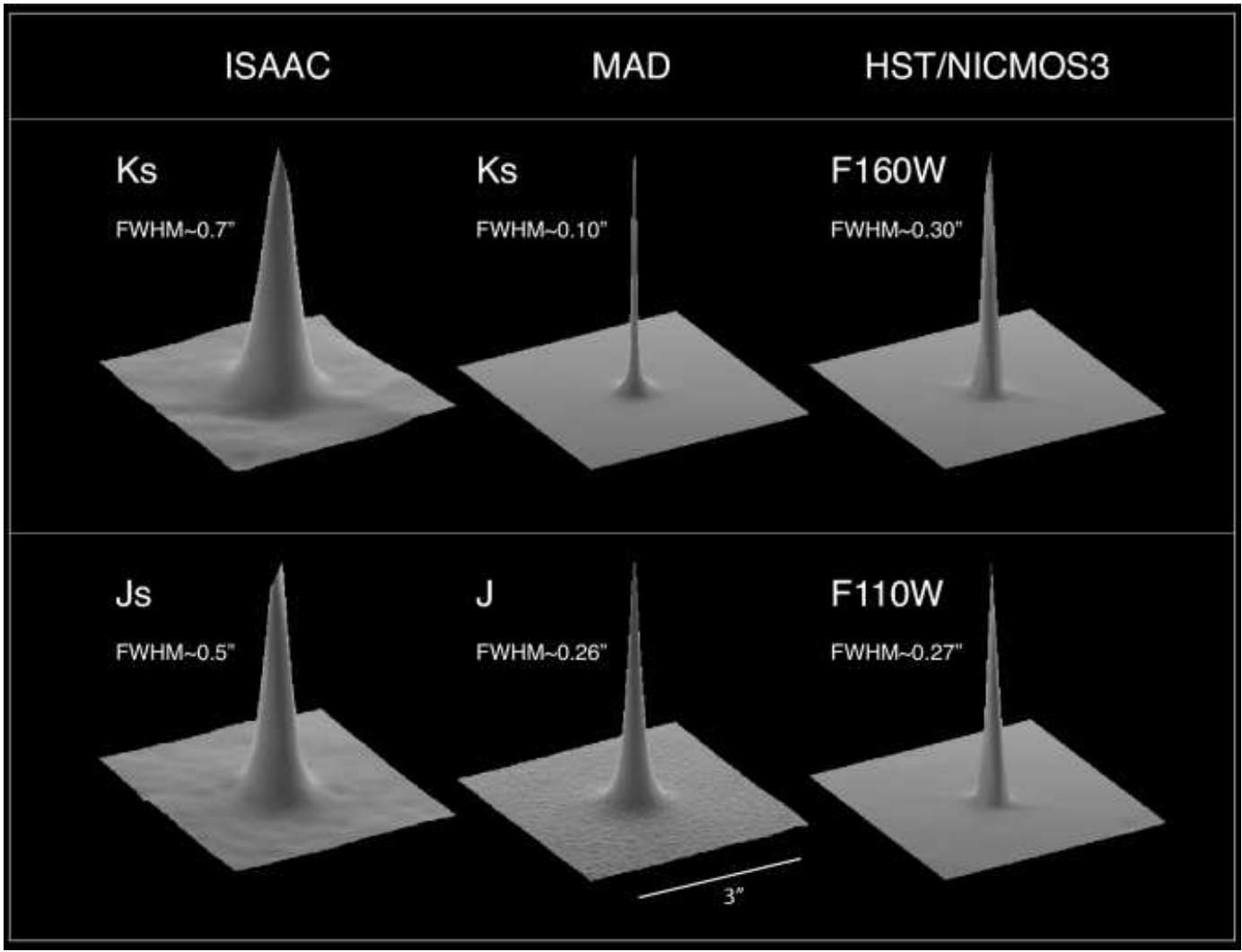}
      \caption{Comparison of the PSF of ISAAC (left, Js and Ks-bands), MAD (center, J and Ks-bands) and HST/NICMOS3 (right, F110W and F160W, no longer wavelength image being available). Even at low strehl in the J band, the resolution of MAD is excellent. The scale and FWHM are indicated. All PSFs have been normalized in peak.}
         \label{psf}
   \end{figure*}

Figure \ref{strehlmap} shows strehl and FWHM maps computed in the Ks band. The under-sampling of the PSF in the J and H bands prevent us to draw similar maps in these bands, although we can expect a similar behavior but with lower correction performances. The strehl ranges from 7.5\% in the edges to 25\% in the center. In the Ks band, the FWHM ranges between 0\farcs09 and 0\farcs15. As expected, the quality of the correction follows closely the geometry of the 3 reference stars. Figure \ref{psf} shows a comparison between the PSF of diffraction-limited images of the Trapezium obtained with HST WFPC2 and NICMOS, VLT MAD, and seeing-limited images obtained with VLT ISAAC, for a Ks=11~mag star. The average resolution of the MAD J and H-band images is 0\farcs250 and 0\farcs180 respectively. The ISAAC images were obtained with a seeing\footnotemark[1]{} ranging between 0\farcs5 and 0\farcs8.

\subsection{Limit of sensitivity}
\begin{figure*}
   \centering
   \includegraphics[width=0.95\textwidth]{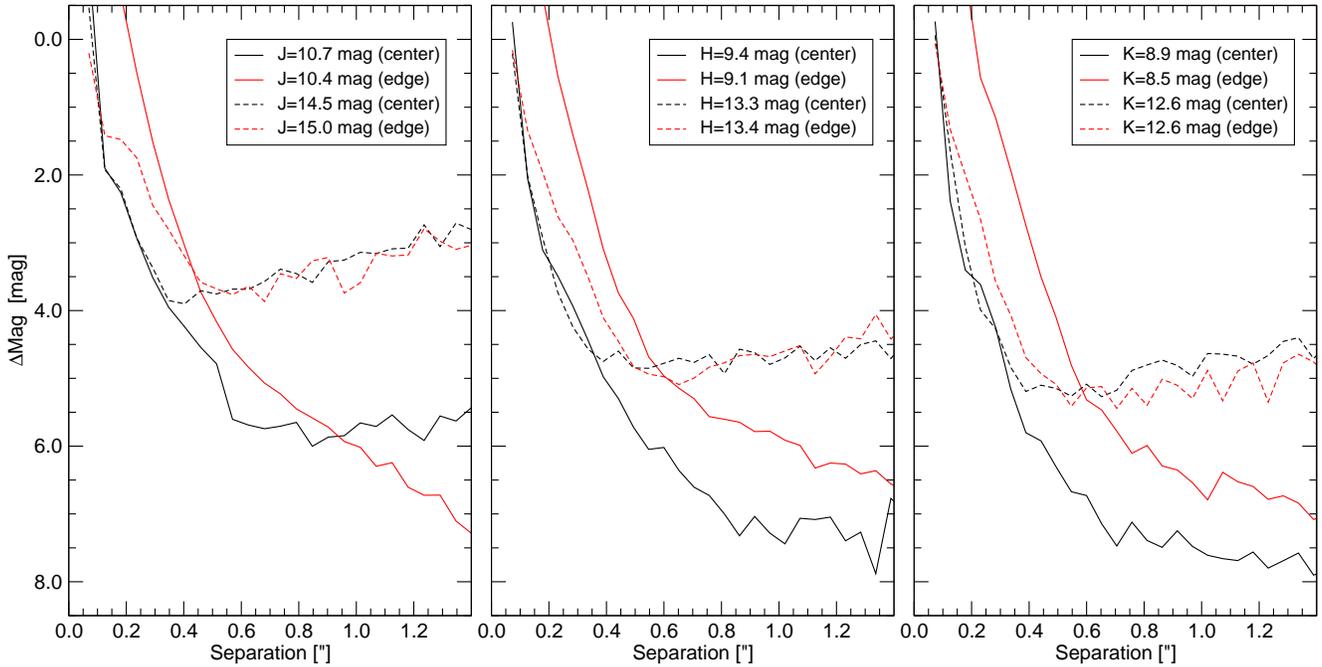}
      \caption{Limit of sensitivity of the MAD J, H and Ks images (respectively left, center and right panels). The limit of sensitivity have been computed as the 3-$\sigma$ noise on the PSF radial profiles for a set of bright (Ks$\approx$9~mag, full lines) and faint (Ks$\approx$12.6~mag, dashed lines) stars, located in in the central area of the image (black lines) where the AO correction is best, and in the outer edges of the image (red lines) where the correction is worst.}
         \label{limsens}
   \end{figure*}

One of the advantages and aims of adaptive optics is to provide high contrast images at close separations. We have computed the limit of sensitivity for stars located in a region of good AO correction  (hereafter referred as center) and in a region of worse AO correction (hereafter referred as edges), for both bright (Ks$\approx$9~mag) and faint objects (Ks$\approx$12.6~mag). Figure \ref{limsens} shows the results. For faint objects, the main limitation at separations greater than 0\farcs4 comes from the high nebular background. In just 5~min and under relatively poor condition (airmass=1.5 and seeing $\approx$1\farcs0), MAD can easily reach $\Delta$Mag=6~mag at 0\farcs6, 0\farcs5 and 0\farcs4 in J, H and Ks respectively on a bright source in the center. In the edges and at 0\farcs6 separation, MAD still reaches $\Delta$Mag=4.8, 5.0 and 5.3~mag in respectively J, H and Ks for bright stars.

\begin{figure*}
   \centering
   \includegraphics[width=0.9\textwidth]{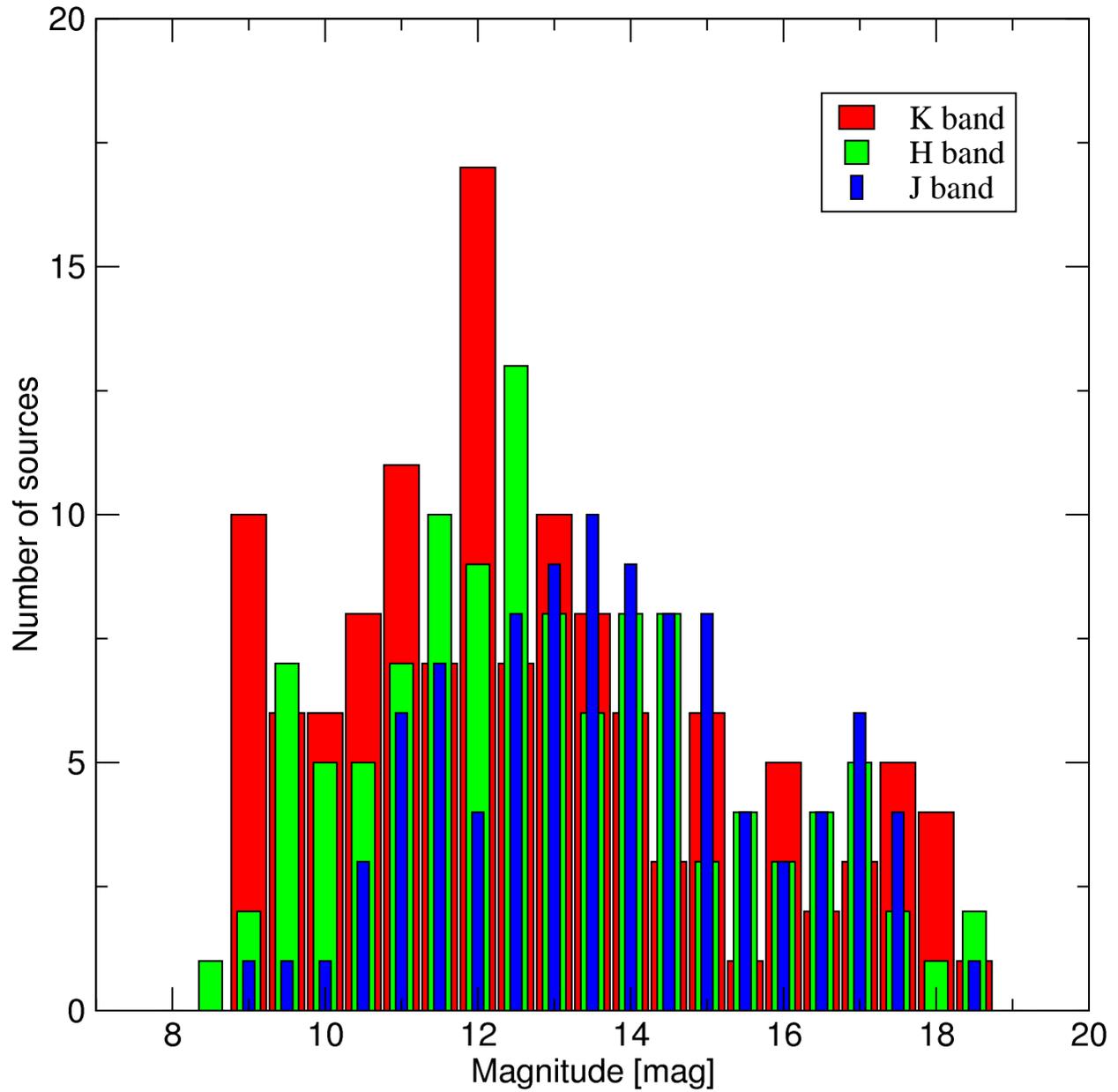}
      \caption{Distribution of magnitude of the sources in the MAD J (blue), H (green) and Ks (red) band images. Because of the strong and variable nebulosity, it is not possible to derive a limit of completeness. The faintest objects detected in the images have Ks=18.1~mag, H=18.2~mag and J=18.5~mag.}
         \label{distrib_mag}
   \end{figure*}

Figure \ref{distrib_mag} shows the distribution of magnitude of the sources detected in the images. Sources brighter than $\approx$8~mag were saturated or above the detector linearity limit. We detect sources as faint as Ks=18.1~mag, H=18.2~mag and J=18.5~mag.

\section{Multiple systems}

With an average resolution of $\approx$0\farcs1 in the Ks band, the MAD images resolve a number of multiple systems. The bright binaries $\theta^{1}$ Orionis A and B are resolved but saturated in all broad band filters. We therefore use the Br$\gamma$ image to derive accurate relative astrometry. We do not resolve the 0\farcs030 companion of $\theta^{1}$ Orionis C, but resolve the faint fourth component near $\theta^{1}$ Orionis B  \citep{1999AJ....117.1375S,2003A&A...402..267S} in the H and Ks band images. The saturation of $\theta^{1}$ Orionis A and B does not allow us to provide a useful relative photometry in the broad band filters. A standard PSF photometry of the B4 companion gives H=11.2 and Ks=10.6~mag, consistent with the results of \citet{1999AJ....117.1375S,2003A&A...402..267S}.  Four additional previously known binaries \cite[TCC-101, TCC-094, TCC-075, and TCC-105, ][]{1994AJ....108.1382M} are resolved in the MAD images. We also confirm the multiplicity of TCC-055, which was suspected to be binary by \citet{1998PhDT........33P} using speckle imaging. We have measured the relative astrometry and photometry using dual-PSF fitting following the method described in \citet{2003AJ....126.1526B} and adapted to MAD. For each objects, we used a set of 3 reference PSF selected near the target in order minimize anisoplanatic and time variability effects. Table \ref{prev_bin} gives a summary of the relative astrometry and photometry measured for the multiple systems. 

\subsection{Previously known multiple systems}

\vspace{0.5cm}

\begin{figure*}
   \centering
   \includegraphics[height=0.4\textheight]{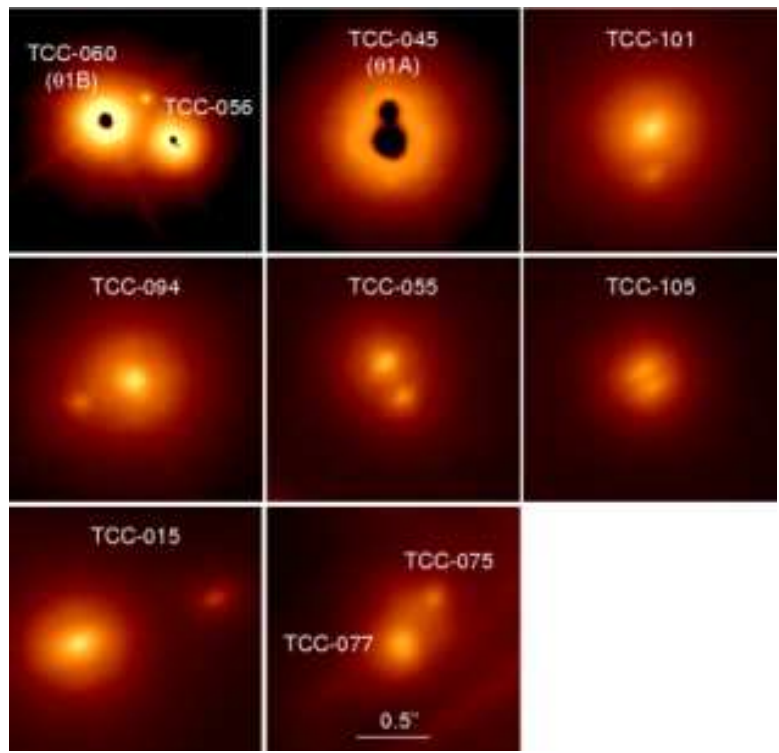}
      \caption{MAD Ks-band stamps of the previously known multiple systems with $\delta<$1\arcsec.  The two bright Trapezium stars $\theta$-Ori B and A are saturated, but their companions are clearly visible. North is up and east is left. The scale is indicated.  The objects are named with their TCC number \citep{1994AJ....108.1382M}.}
         \label{prev_bin_fig}
   \end{figure*}

\noindent \emph{TCC-094 --} has first been resolved as a double by \citet{1994ApJ...421..517P} using HST. They report a separation of 0\farcs371, $\Delta$I=2.52~mag and $\Delta$V=1.94~mag. The binary was resolved  again in December 1996 by \citet{1999AJ....117.1375S} using adaptive optics with a separation of 0\farcs298 and $\Delta$Ks=0.00~mag. The separation in the MAD images (0\farcs384) is significantly larger than that reported by \citet{1999AJ....117.1375S}. We measure $\Delta$Ks=3.19~mag, consistent with the optical relative photometry measured with HST, but inconsistent with the Ks-band relative photometry reported by \citet{1999AJ....117.1375S}. Figure \ref{prev_bin_fig} shows clearly that the flux ratio is far from being equal to unity, as reported by \citet{1999AJ....117.1375S}.  
A possible explanation is that one or both components of the binary are photometrically variable, which would not be surprising considering their young age. In that case, \citet{1999AJ....117.1375S} would have been observing the sources during a strong outburst of the faint component. This is further supported by the significant variability of the individual components. \citet{1999AJ....117.1375S} report Ks=9.90~mag for the primary and the secondary, while we measure Ks=8.5~mag for the primary and Ks=11.7~mag for the secondary.

\vspace{0.5cm}

\noindent \emph{TCC-015 --} has first been resolved as a binary by \citet{1994ApJ...421..517P} using HST. They report a separation of 1\farcs013, $\Delta$I=1.45~mag and $\Delta$V=1.06~mag, in good agreement with our new measurements.

\vspace{0.5cm}

\noindent \emph{TCC-105 --} has first been reported as a binary by \citet{1994ApJ...421..517P} using HST. They report a separation of 0\farcs125, $\Delta$I=0.25~mag and $\Delta$V=0.81~mag, in good agreement with our new measurements. We note that the companion is very much redder than its primary, as shown in Figure \ref{cmddiag}, suggesting that it is significantly cooler.

\vspace{0.5cm}

\noindent \emph{TCC-077/TCC-075 --} has been resolved in November 1994 by \citet{1998ApJ...500..825P} with a separation of 0\farcs33$\pm$0\farcs02 and $\Delta$Ks=1.43~mag.  It was later classified as a binary proplyd by \citet{1996AJ....111..846O} using HST. The object is clearly resolved in the MCAO images, and the PSF of the companion is consistent with that of a stellar object.

\vspace{0.5cm}

\noindent \emph{TCC-101 --} has been resolved in November 1994 by \citet{1998ApJ...500..825P} using speckle interferometry. They report a separation of 0\farcs22$\pm$0\farcs01 and P.A=121\degr. Thirteen years later, we report a significantly larger separation and smaller P.A. The proper motion of the cluster being mostly along the line of sight, the difference likely corresponds to orbital motion. 

\subsection{Confirmation of a new multiple system}

The deep near-IR images allow us to resolve TCC-055 as a new visual binary. The object was reported as a possible binary in the speckle observations of \citet{1998PhDT........33P}, but discarded in their final sample of multiple systems \citep{1998ApJ...500..825P}. Figure \ref{prev_bin_fig} shows that the pair is nicely and unambiguously resolved in the new MAD images. TCC-055 is not resolved in the optical HST WFPC2 \citep{1996AJ....111..846O} and ACS images \citep{2005sfet.confE..44R}. The luminosity and colors of the companion (Ks=12.21~mag, H-Ks=2.99~mag) make it a good very low mass candidate. The companion could also be an unrelated background source, which would explain its very red colors and non-detection in the optical. Simple statistical considerations on the density of the cluster and the high extinction make the probability to find an unrelated background source in such a field relatively low \citep{1998ApJ...500..825P}.

\section{A new census of Trapezium sources} 
Using the unprecedented depth and resolution of the MAD images, we have detected a total of 128 sources within the 1\arcmin$\times$1\arcmin\, FoV of the Ks image. We compare this sample with the catalogues reported in previous studies.
 
\subsection{Near-IR detection of faint objects}
The deep L'-band ISAAC images of \citet{2004AJ....128.1254L} produced a catalogue of a previously unrecognized population of strongly reddened objects in the Trapezium cluster. The total area covered by their survey is much larger than that covered by the MAD images, and in the following discussion we will refer only to the 106 sources reported in their catalogue and located in the field of view of the MAD images.

Even though deeply embedded sources are difficult to detect at shorter wavelengths, all 106 ISAAC L' sources but one are detected in the J, H and Ks image. The one source undetected in the MAD images, reported as number LMLA2004~169 in their catalogue, is located in a region of extreme extinction. 

A total of two objects identified in their L' images were lacking previous J, H, Ks detection. One is detected in the MAD Ks image (number 179 in their catalogue, CACAO-2 in Table \ref{new_sources}). A total of 3 previously known near-IR sources were not detected in their L' images, and all 3 are detected in the MAD images.

\subsection{X-ray survey}
The \emph{Chandra Orion Ultradeep Project} \citep[hereafter COUP, ][]{2005ApJS..160..319G} has produced a very deep catalogue of X-ray sources in Orion. All COUP sources located in the field of the MAD images but 7 are detected in the Ks-band image, the missing ones being COUP~J053517.9-052326, COUP~J053517.8-052321.6, COUP~J053517.7-052320.1, COUP~J053517.0-052302.9, COUP~J053516.3-052301.4, COUP~J0535160-052334.4, and COUP~J053514.5-052315.9. 

 \subsection{New sources}
In addition to the previously known X-ray and near-IR sources mentioned above, we report the detection of 12 new sources in the Ks images. Four have H band counterparts, with H-K$\approx$1~mag for each of them. One source also has a J band counterpart. Table \ref{new_sources} shows the properties of these sources. Two of these sources (CACAO-11 and -12) lie close to detector artefacts and are suspicious, even though their PSFs are consistent with point sources. These two sources require further investigation, and we do not include them in the rest of the discussions in this article. One source (CACAO-2) was previously detected in the L' band \citep{2004AJ....128.1254L}, but not in the Ks band. These new detections shows that the previous surveys have not reached the bottom of the luminosity function of the Trapezium cluster, and that more sources are likely to be discovered. A more detailed study of the IMF using the MAD catalogue will be presented in Barrado y Navascu\'es et al. (in prep.).

\begin{center}
\begin{deluxetable}{lccccccc}
\tablecaption{New near-IR detections \label{new_sources}}
\tablewidth{0pt}
\tablehead{
\colhead{ID\tablenotemark{1}} & \colhead{R.A}      & \colhead{Dec}    & \colhead{J}  & \colhead{H}     & \colhead{Ks} \\
\colhead{}       & \colhead{(J2000)} & \colhead{(J2000)} & \colhead{[mag]} & \colhead{[mag]} & \colhead{[mag]} 
}
\startdata
CACAO-1  & 05:35:14.5  & -05:23:41.7    & \nodata & \nodata & 17.0  \\
CACAO-2\tablenotemark{2}   & 05:35:14.7  & -05:23:30.3    & \nodata & \nodata & 17.2  \\
CACAO-3  & 05:35:15.2  & -05:23:39.2    & \nodata & \nodata & 17.9  \\
CACAO-4  & 05:35:15.5  & -05:23:18.2    & \nodata & \nodata & 17.4  \\
CACAO-5  & 05:35:15.8  & -05:23:40.2    & \nodata & \nodata & 17.8  \\
CACAO-6 & 05:35:16.0  & -05:23:30.7    & \nodata & 17.9    & 16.9  \\
CACAO-7 & 05:35:16.1  & -05:23:11.0    & \nodata & 17.2    & 16.1  \\
CACAO-8  & 05:35:16.9  & -05:22:45.3    & \nodata & \nodata & 17.2  \\
CACAO-9 & 05:35:16.9  & -05:23:38.1    & 17.1    & 16.5    & 15.7  \\
CACAO-10  & 05:35:17.0  & -05:23:08.9    & \nodata & \nodata & 17.9  \\
CACAO-11\tablenotemark{3}  & 05:35:16.7  & -05:22:45.3    & \nodata & 18.1    & 17.0  \\
CACAO-12\tablenotemark{3}  & 05:35:16.7  & -05:23:30.1    & \nodata & \nodata & 17.5  \\
\enddata
\tablenotetext{1}{CACAO stands for {\bf\emph{CA}}ndidates from multi-{\bf\emph{C}}onjugate {\bf \emph{A}}daptive {\bf\emph{O}}ptics}
\tablenotetext{2}{Discovered in the L'-band only by \citet{2004AJ....128.1254L}}
\tablenotetext{3}{Close to detector artifact.}
\end{deluxetable}
\end{center}

\section{Extended objects}
A number of extended sources are resolved in the MAD images. Two objects were resolved in the ISAAC L' images \citep[TCC-065 and TCC-086, respectively LMLA~268 and LMLA~309 in ][]{2004AJ....128.1254L}, but their proximity to the bright Trapezium B stars does not allow us to resolve them in the MAD images.

\subsection{Proplyds}
\begin{figure*}
   \centering
   \includegraphics[height=0.5\textheight]{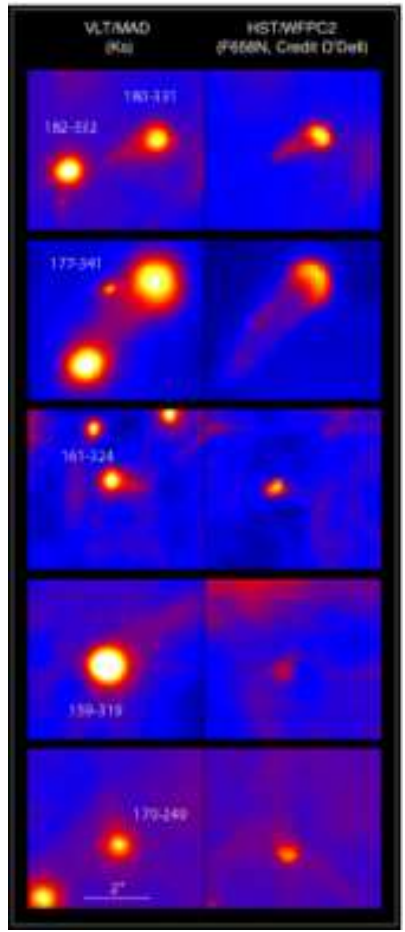}
      \caption{MAD Ks-band stamps of some spatially resolved proplyds (left), and corresponding HST/WFPC2 F658N image from \citet{1996AJ....111..846O}. The name of the proplyds \citep[following ][denomination]{1996AJ....111..846O} and the scale are indicated. North is up and east is left.}
         \label{proplyds}
   \end{figure*}

All the sources associated to proplyds \citep[][and references therein]{2005A&A...441..195V} located in the FoV of the MAD images are detected, but only some of them are resolved in the Ks image. Figure \ref{proplyds} shows examples of resolved proplyds.

\subsection{HC~419 \label{blobby}}

\begin{figure*}
   \centering
   \includegraphics[width=0.95\textwidth]{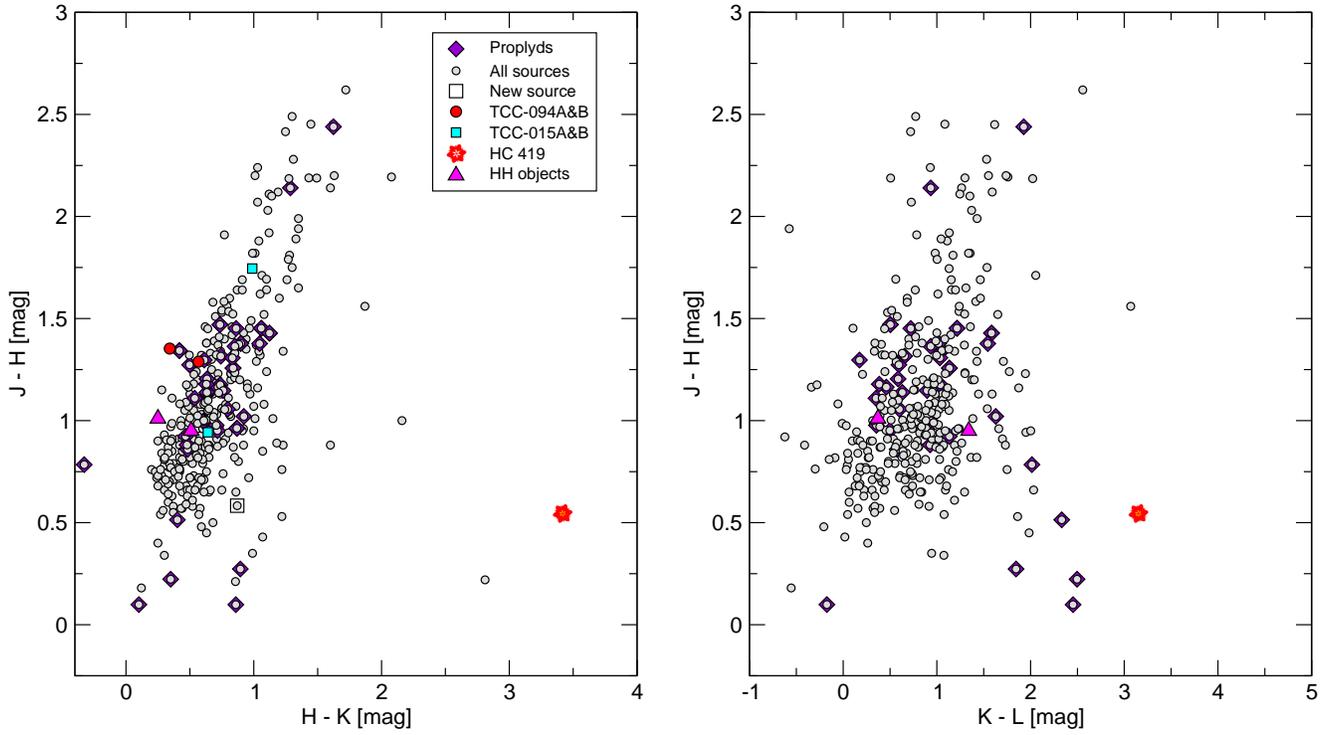}
      \caption{\emph{Left Panel:}(J-H) vs (H-Ks) color-color diagram of the combined \citet{2004AJ....128.1254L} and MAD samples (grey dots). For the sources within the MAD FoV, we plot the MAD J, H, Ks photometry and the cross-matched L' photometry of \citet{2004AJ....128.1254L}. For the objects outside the MAD images, we plot the J, H, Ks, L' photometry of \citet{2004AJ....128.1254L}. The components of the two binaries with J, H and Ks detections, TCC-094 and TCC-015 (see Table \ref{prev_bin}) are indicated with red dots and blue squares respectively. The new source CACAO-9 with J, H and K detection is represented with an open square. Two HH objects, HH~513 and HH~562 \citep{2001AJ....122.2662O}, are overplotted for comparison (magenta triangles). The very red object HC~419 is represented with a red flower. Known proplyds are over-plotted with purple squares. \emph{Right Panel:} (J-H) vs (Ks-L') color-color-diagram of \citet{2004AJ....128.1254L}. The same symbols are used.}
         \label{cmddiag}
   \end{figure*}

Figure \ref{cmddiag} shows (J-H) vs (H-Ks) and (J-H) vs (Ks-L') color-color diagrams of the MAD sources with J, H, Ks detection and L' band counterparts in \citet{2004AJ....128.1254L} catalogue, as well as all the objects of \citet{2004AJ....128.1254L} outside the MAD FoV. One object has (H-Ks) and (Ks-L') colors clearly much redder than any other object in the sample, and was included by \citet{2004AJ....128.1254L} in their list of  deeply embedded objects because of its very red colors. It has the sequence number 214 in their catalogue. The source was first reported by \citet{2000ApJ...540..236H} in their Keck H and K-band survey, and is identified with the sequence number 419 in their catalogue. We hereafter refer to it as HC~419 ($\alpha$=05h35m15.5s, $\delta$=-05\degr22\arcmin46.5\arcsec, J2000). It is also reported in \citet{2002ApJ...573..366M} (sequence number 543), but without any photometric measurements. Its K-L'=3.22~mag is surpassed only by one object, TPSC-78 \citep[K-L=4.22~mag, ][]{2000AJ....120.3162L}. HC~419 was subsequently reported by several authors: \citet{2005AJ....129.1534R} detected it in the N-band (identified as MAX-84 in their catalogue), \citet{2005AJ....130.1763S} obtained 11.7~$\mu$m photometry, and it has a counterpart in the COUP survey, identified as COUP~J053515.5-052246. It was not detected in the published HST optical images, but was reported (unresolved) in the HST NICMOS F160W image presented in \citet{2000ApJ...540.1016L} (sequence number 108 in their catalog). A careful inspection of the images shows that HC~419 was detected in the unpublished ISAAC images. 

We have searched the VLT public archive for mid-IR data and found that HC~419 fell in the edge of the field of view of observations performed during a Science Verification program (program 60.A-9263, P.I. Lagage). We retrieved the data and processed them using a customary pipeline based on IDL routines. The pipeline includes bad pixel correction and the shift-and-add of the individual frames, each of them corresponding to a chopping position \citep{2005Msngr.119...25}. The target is detected (but unresolved) with a signal-to-noise ratio of $\approx$10 in the 20~min PAH1 exposure only ($\lambda_{\rm cen}=$8.59~$\mu$m, $\Delta\lambda=$0.42~$\mu$m). It is not detected in the \ion{Ar}{iii} ($\lambda_{\rm cen}=$8.99~$\mu$m, $\Delta\lambda=$0.14~$\mu$m), and we derive a 3-$\sigma$ upper limit on the flux. 

\begin{figure*}
   \centering
   \includegraphics[width=0.55\textwidth]{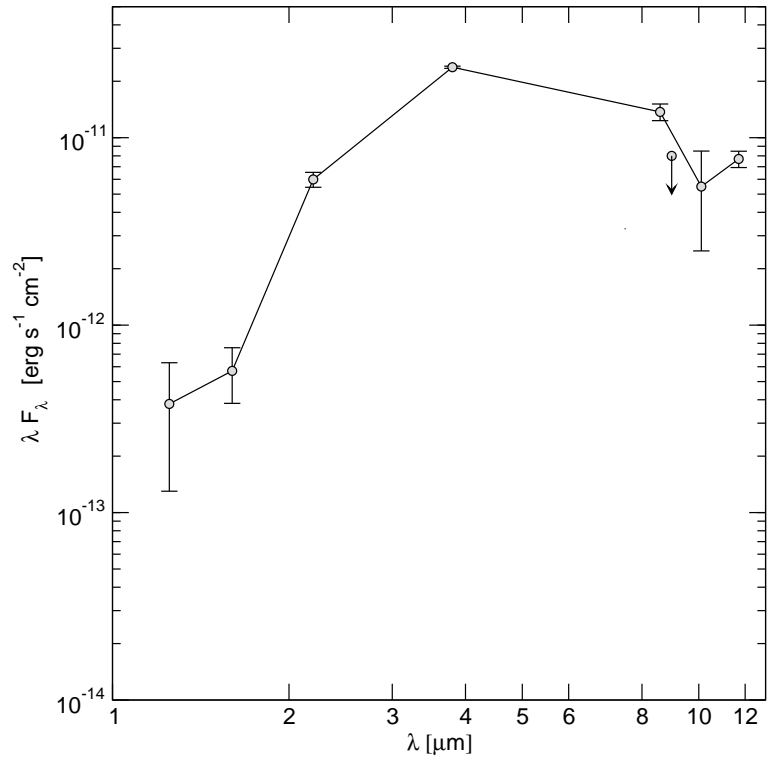}
      \caption{Spectral Energy Distribution of HC~419. The slope at wavelength longer than 2.2~$\mu$m is characteristic of intermediate Class~I--II objects.}
         \label{sed}
   \end{figure*}

Table \ref{blob} gives an summary of the photometry of HC~419 and Figure \ref{sed} shows its spectral energy distribution (SED). 

\begin{figure*}
   \centering
   \includegraphics[width=0.95\textwidth]{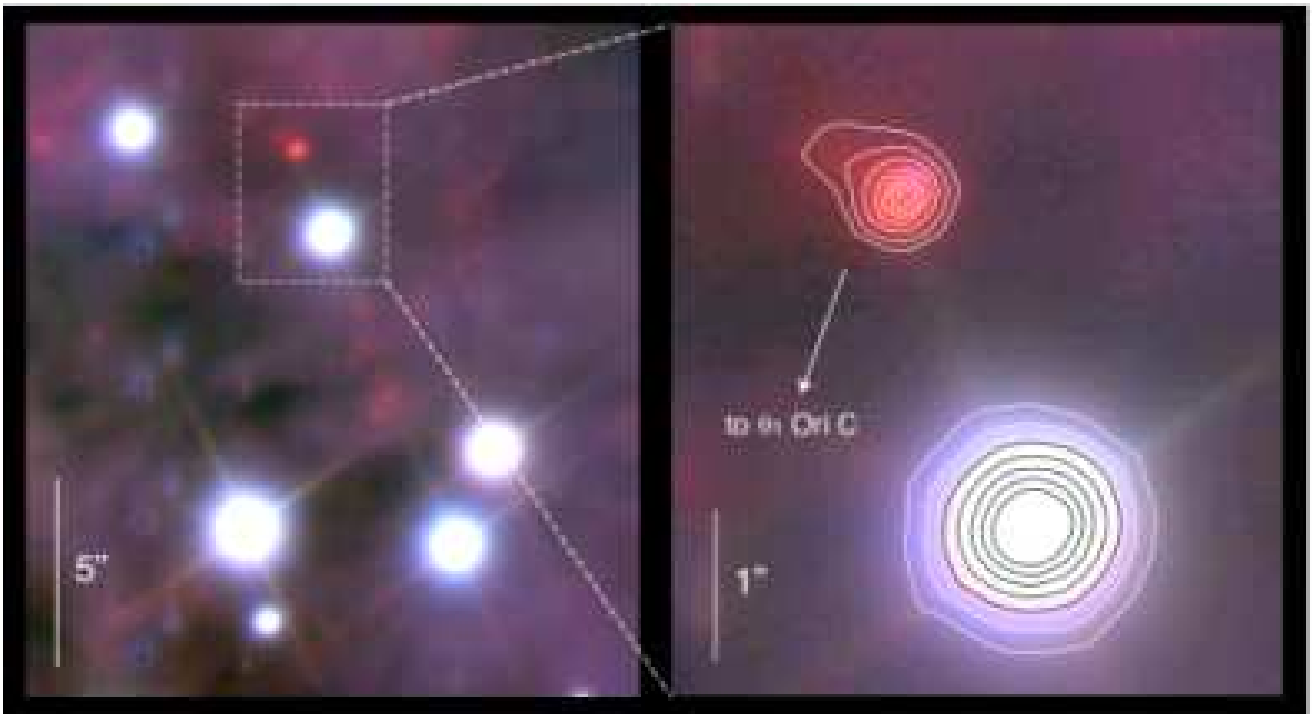}
      \caption{MAD 3-color stamps around HC~419, the reddest object of the MAD sample. The right panel is a zoom in of the left panel with contour plots showing the extension of the object when compared to an unresolved neighboring stellar source. It is associated with nebulosities. North is up and east is left, and the scale is indicated. See also Fig. \ref{2mass}.}
         \label{newproplyd}
   \end{figure*}

A careful inspection of the MAD images shows that this object is elongated in the Ks band only, and lies in a region of high red nebulosity (see Fig. \ref{newproplyd}). The color and spatial extension of this source make it a very interesting object, and the new MAD images provide a new insight on its nature. Both its unusual colors and the direction of elongation, which is almost orthogonal to the direction of $\theta^{1}$ Ori C (see Fig. \ref{newproplyd}), allow us to almost certainly rule out the eventual proplyd nature of this extended object. Figure \ref{cmddiag} shows that the colors are also inconsistent with Herbig Haro objects.

The larger column density derived by the COUP team for that source compared to its closest neighbors confirms that the object must be very deeply embedded. For HC~419, they report $\log{n_{\rm H}}=$22.72$\pm$0.06~cm$^{-2}$. The value reported for the two closest neighbors (2\farcs3 and 4\farcs4 away, see Figure \ref{newproplyd}) are smaller by a factor $>$10, with $\log{n_{\rm H}}$=21.68$\pm$0.02~cm$^{-2}$ and 21.47$\pm$0.08~cm$^{-2}$ respectively. This is further supported by the clear elongation observed in the Ks band which  suggest that the unusual colors are not due to foreground extinction purely related to the molecular cloud only, but also in part to extended emission associated to the object itself such as e.g a dense envelope and/or a circumstellar disk, making HC~419 likely protostellar in nature. 

The slope of the SED between the H and 11.7~$\mu$m fluxes provides an additional evidence of the protostellar nature  of HC~419. In the morphological classification scheme of SEDs devised by \citet{1987...Lada}, HC~419 resemble to an intermediate Class~I--II star, with a relatively flat SED at wavelengths longer than 2.2~$\mu$m, and a spectral index $a=d\log{\lambda\, \mathcal{F}_{\lambda}}/d\log{\lambda}$ in the 2.2--11.7~$\mu$m range of $a$=-0.1 and $a=$-1.1 in the range 3.8--11.7~$\mu$m. 

\begin{center}
\begin{deluxetable}{lcc}
\tablecaption{Photometry of HC~419 \label{blob}}
\tablewidth{0pt}
\tablehead{
\colhead{$\lambda$} & \colhead{Flux}    & \colhead{Ref.}  \\
\colhead{[$\mu$m]}          & \colhead{[mJy]}   & \colhead{} 
}
\startdata
1.25  & 0.3$\pm$0.1    & (1)   \\
1.6   & 0.3$\pm$0.1    & (1) \& (5)  \\
2.2   & 4.4$\pm$0.4    & (1) \& (5)  \\
3.8   & 30.1$\pm$0.4   & (2)   \\
8.49  & 39.3$\pm$4.0   & (1)  \\
8.99  & $<$24 (3-$\sigma$)          & (1)  \\
10.1  & 18.5$\pm$9.75  & (3)   \\
11.7  & 30$\pm$3       & (4)   \\

\enddata
\tablerefs{(1) This work; 
(2) \citet{2004AJ....128.1254L}; 
(3) \citet{2005AJ....129.1534R}; 
(4) \citet{2005AJ....130.1763S}; 
(5) \citet{2005ApJS..160..319G} }
\end{deluxetable}
\end{center}

\section{Conclusion and future prospects}
Originally designed as a laboratory experiment, MAD has delivered images on the sky of a quality comparable to the most competitive instruments available nowadays. For a fraction of the cost, it provides a resolution equivalent or better than HST NICMOS, with both the sensitivity and field of view of an 8~m class near-IR instrument such as ISAAC. In this paper, we have shown that MAD not only successfully demonstrates the technical feasibility of MCAO, but also shows that a large variety of scientific results will be accessible thanks to this new versatile technology. With classical AO, studies requiring high spatial resolution, high sensitivity, and high contrast are limited to relatively small fields. MCAO will extend these studies to arcminute size fields. MCAO will certainly play a key role in the current context of large scale surveys and ELT development. 



\begin{acknowledgements}
The authors thank Gaspard Duch\^ene, Silvia Vicente, F. Marchis and J.-L. Monin for interesting discussions and comments about this work.
H. Bouy acknowledges the funding from the European Commission's Sixth Framework Program as a Marie Curie Outgoing International Fellow (MOIF-CT-2005-8389). Partial financial support was provided by the spanish Ministerio de Educacion y Ciencia project AYA2006-12612. This work is based on observations obtained with the MCAO Demonstrator (MAD) at the VLT (ESO Public data release) which is operated by the European Southern Observatory. The MAD project is led and developed by ESO with the collaboration of the INAF-Osservatorio Astronomico di Padova (INAF-OAPD) and the Faculdade de Ci\^encias de Universidade de Lisboa (FCUL). This work is based on observations made with ESO Telescopes at the La Silla or Paranal Observatories under programme 66.C-0294. This publication makes use of data products from the Two Micron All Sky Survey, which is a joint project of the University of Massachusetts and the Infrared Processing and Analysis Center/California Institute of Technology, funded by the National Aeronautics and Space Administration and the National Science Foundation.  This makes use of observations made with the NASA/ESA Hubble Space Telescope, obtained from the public data archive at the Space Telescope Institute. STScI is operated by the association of Universities for Research in Astronomy, Inc. under the NASA contract  NAS 5-26555. This work has made use of the Vizier Service provided by the Centre de Donn\'ees Astronomiques de Strasbourg, France \citep{Vizier}. This research used the facilities of the Canadian Astronomy Data Centre operated by the National Research Council of Canada with the support of the Canadian Space Agency. 

\end{acknowledgements}

\begin{center}
\begin{deluxetable}{lccccccc}
\tablecaption{New relative astrometry and photometry of previously known binaries \label{prev_bin}}
\tablewidth{0pt}
\tablehead{
\colhead{Object} & \colhead{R.A}      & \colhead{Dec}    & \colhead{$\delta$}  & \colhead{P.A}     & \colhead{$\Delta$K} & \colhead{$\Delta$H} & \colhead{$\Delta$J} \\
\colhead{}       & \colhead{(J2000)} & \colhead{(J2000)} & \colhead{[\arcsec]} & \colhead{[\degr]} & \colhead{[mag]} & \colhead{[mag]} & \colhead{[mag]}
}
\startdata
TCC-094	         & 05:35:17.1         & -05:22:50.0      & 0.384$\pm$0.004     & 248.1$\pm$0.3      & 3.19$\pm$0.10 & 3.2$\pm$0.3 & 2.1$\pm$0.3 \\
TCC-055		 & 05:35:16.1         & -05:22:54.0      & 0.256$\pm$0.004     & 153.1$\pm$0.3      & 1.26$\pm$0.10 & 2.2$\pm$0.3 & \nodata \\
TCC-105              & 05:35:18.0         & -05:23:35.4      & 0.134$\pm$0.004     & 152.2$\pm$0.3      & 0.41$\pm$0.10 & 0.5$\pm$0.3 & \nodata \\
TCC-101	         & 05:35:17.8         & -05:23:15.6      & 0.303$\pm$0.004     & 178.3$\pm$0.3      & 3.31$\pm$0.10 & \nodata & \nodata \\
TCC-015              & 05:35:14.8         & -05:23:04.6      & 1.022$\pm$0.004     & 288.2$\pm$0.3      & 3.65$\pm$0.10 & 3.3$\pm$0.3 & 2.5$\pm$0.3 \\
TCC-077/075          & 05:35:16.8        & -05:23:26.2      & 0.396$\pm$0.004     & 34.6$\pm$0.3       & 1.30$\pm$0.10 & 1.2$\pm$0.3 & \nodata \\
$\theta^{1}$ Ori A1/A2  & 05:35:15.8 & -05:23:14.3 &  0.201$\pm$0.004 & 4.3$\pm$0.3  & \nodata & \nodata & \nodata \\
$\theta^{1}$ Ori B1/B2  & 05:35:16.1 & -05:23:07.3 & 0.930$\pm$0.004 & 209.4$\pm$0.3 &  \nodata & \nodata & \nodata \\
$\theta^{1}$ Ori B1/B4  &            &             & 0.593$\pm$0.004 & 298.2$\pm$0.3  &  \nodata & \nodata & \nodata \\
$\theta^{1}$ Ori B1/B3  &            &             & 1.029$\pm$0.004 & 250.7$\pm$0.3  &  \nodata & \nodata & \nodata \\
$\theta^{1}$ Ori B2/B3  & 05:35:16.1 & -05:23:07.1   & 0.117$\pm$0.004 & 219.7$\pm$0.3  &  \nodata & \nodata & \nodata \\
\enddata
\tablecomments{TCC ID from \citet{1994AJ....108.1382M}. $\theta^{1}$ Ori A and B components were saturated in the broad band images, and the measurements were made in the Br$\gamma$ image.}
\end{deluxetable}
\end{center}

\end{document}